# THE COLLABORATIVE EFFECTS OF INTRINSIC AND EXTRINSIC IMPURITIES IN LOW RRR SRF CAVITIES[*]

K. Howard[†], Y.-K. Kim, University of Chicago, Chicago, IL, USA
D. Bafia, A. Grassellino, Fermi National Accelerator Laboratory, Batavia, IL, USA

## Abstract

The superconducting radio-frequency (SRF) community has shown that introducing certain impurities into high-purity niobium can improve quality factors and accelerating gradients. We question why some impurities improve RF performance while others hinder it. The purpose of this study is to characterize the impurity profile of niobium with a low residual resistance ratio (RRR) and correlate these impurities with the RF performance of low RRR cavities so that the mechanism of impurity-based improvements can be better understood and improved upon. The combination of RF testing and material analysis reveals a microscopic picture of why low RRR cavities experience low temperature-dependent BCS resistance behavior more prominently than their high RRR counterparts. We performed surface treatments, low temperature baking and nitrogen-doping, on low RRR cavities to evaluate how the intentional addition of oxygen and nitrogen to the RF layer further improves performance through changes in the mean free path and impurity profile. The results of this study have the potential to unlock a new understanding on SRF materials and enable the next generation of SRF surface treatments.

## INTRODUCTION

As we approach the theoretical limit of Nb for superconducting radio-frequency (SRF) cavities, the last decade has brought immense improvements in quality factor ($Q_0$) and accelerating gradients though intentionally added impurities into the Nb surface [1, 2]. Many SRF studies follow a "clean bulk dirty surface" technique to optimize the BCS resistance ($R_{BCS}$) by adding extrinsic impurities to the surface layer of high purity Nb [3–5]. Advancements have been made with N through N-doping, where cavities experience an anti-$Q_0$ slope and record breaking $Q_0$'s at mid fields [6–8]. O added through a low temperature bake (LTB) has also provided high $Q_0$'s and mitigation of the high field $Q_0$ slope typically seen in electropolished (EP) Nb cavities [9, 10].

The success of intentionally added impurities to the Nb surface has drawn deeper questions about how these impurities affect cavity behavior, and has prompted an investigation of cavities with a low residual resistance ratio (RRR). Low purity Nb has been studied in the past for the purpose of cost reduction and possible high $Q_0$ [11]. In this study, we look to use the intrinsic impurities as a resource to optimize the $R_{BCS}$ and understand the mechanism of impurity-based improvements. We ask how the intrinsic impurities can improve performance, as we observe in extrinsic impurities.

In this study, we investigate a single-cell TESLA-shaped 1.3 GHz cavity with RRR 61. First, the cavity receives EP treatment to make the surface layer and bulk uniform [12]. We measure $Q_0$ versus gradient at 2 K and low temperature (< 1.5 K) in the vertical test stand [2]. The surface resistance is the geometry factor of the cavity divided by the $Q_0$; this can be broken down into the residual resistance ($R_{res}$) and $R_{BCS}$. We compare the performance to its high RRR counterpart in EP condition to understand how the intrinsic impurities affect the bulk and surface behavior of the cavity. We perform a LTB at 120 °C for 48 hours and repeat the testing to evaluate how the addition of the surface oxide to the RF layer further affects performance. We additionally investigate the effect of adding N to the dirty bulk by performing N-doping with the standard 2/6 + 5 $\mu$m recipe [13]. Since the last report [14], we performed secondary ion mass spectrometry (SIMS) on low and high RRR coupons in EP, LTB, and N-doped conditions to characterize their impurity profiles.

## RESULTS

### Quality Factor

We measure the $Q_0$ at a given gradient by maintaining the cavity at its resonant frequency, inputting power via antenna, and then measuring the reflected and transmitted power [15]. The $Q_0$ is the ratio of the energy gain per RF period and dissipated power. The measurements of $Q_0$ at 2 K are graphed in Fig. 1. In general, the $Q_0$'s of the low RRR tests are lower than their high RRR counterparts.

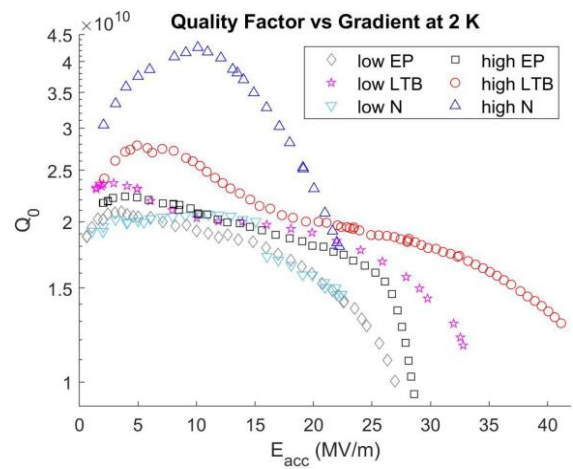

Figure 1: Quality factor at 2 K versus accelerating gradient for EP, LTB, and N-doping on low and high RRR.

---

[*] This manuscript has been authored by Fermi Research Alliance, LLC under Contract No. DE-AC02-07CH11359 with the U.S. Department of Energy, Office of Science, Office of High Energy Physics. This work was supported by the University of Chicago.
[†] khoward99@uchicago.edu

O improves performance of low RRR cavity but with a weaker response than we see in high RRR cavities, as the LTB treatment delays $Q_0$ slope in low RRR less than in high RRR. The low RRR cavity did not show a strong high field $Q_0$ slope in EP condition, so the transition to LTB was not as drastic. The weakened $Q_0$ slope suggests that the intrinsic impurities may capture the free H which is thought to exacerbate the high field $Q_0$ slope [16, 17]. The performance after N-doping is quite similar to EP. N-doping the low RRR cavity did not improve the $Q_0$, unlike high RRR N-doped cavities. The N-doped cavity experienced multi-pacting quenches above 16 MV/m, which trapped magnetic flux and worsened the performance up to its ultimate quench at 22 MV/m.

*Residual Resistance*

The $R_{res}$ taken at low T is temperature-independent, and comes from impurities in the superconducting lattice as well as any trapped flux. The $R_{res}$ measurements are shown in Fig. 2. We observe a significant offset in $R_{res}$ between low and high RRR for all surface treatments, especially at mid gradient. This may suggest that the oxide structure of the low RRR cavity is different or that the intrinsic impurities may drive additional losses.

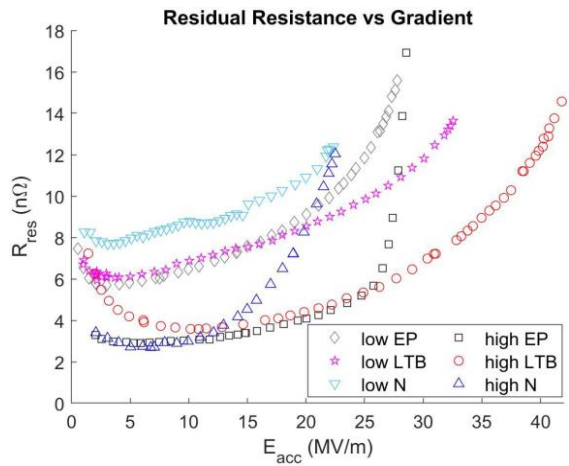

Figure 2: Residual resistance (at low T) versus accelerating gradient for low and high RRR.

The low RRR EP and LTB curves are equal at low and mid gradients. The addition of O to the RF layer did not increase the resistive effect of the intrinsic impurities in the material, and at high gradients the O enables lower $R_{res}$. The low RRR N-doped $R_{res}$ is slightly higher than the corresponding EP and LTB curves. Because N-doping introduces impurities further into the bulk than LTB, it is possible this caused the increase in $R_{res}$. Another possible cause is flux trapped through multipacting quenches during the 2 K test.

*BCS Resistance*

The $R_{BCS}$ is calculated by taking the difference between the total surface resistance at 2 K and low T. This temperature-dependent component of the resistance is caused by the breakdown of cooper pairs with increasing temperature [3, 13]. In Fig. 3, we highlight the low $R_{BCS}$ behavior of the low RRR cavity.

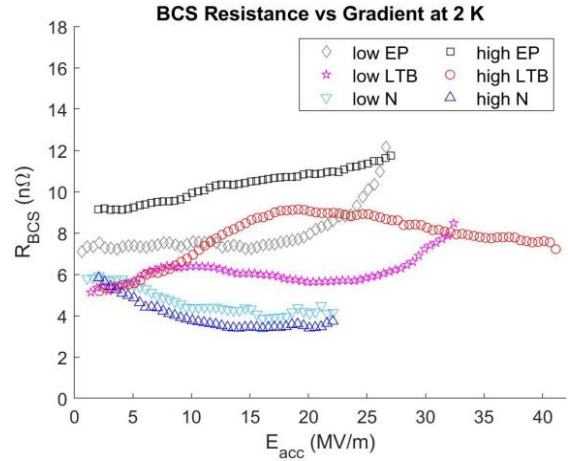

Figure 3: BCS resistance versus accelerating gradient at 2 K for low and high RRR.

The low RRR EP and LTB $R_{BCS}$ are always less than or equal to that of their high RRR counterparts. This benefit is most prominent at mid gradients and lost at high gradients. The high and low RRR LTB curves show a similar behavior of a local maximum and then decrease. It is promising that the LTB lowered the $R_{BCS}$ at all gradients from the EP test. The N-doped test of the low RRR cavity showed similar $R_{BCS}$ than that of the high RRR, but significantly reduced from the EP and LTB tests. N-doping did show improvement from the EP and LTB tests, but it is surprising that the low RRR $R_{BCS}$ is not lower than its high RRR counterpart.

*Impurity Profiles*

The SIMS data is measured as the intensity of each ion versus sputtering time. The impurity profiles shown in Figs. 4, 5, 6, and 7 are the most relevant ions found showing the differences between the surface treatments in low and high RRR. The x axes are normalized by the noise floor of the $Nb_2O_5$ signal at 10 counts of intensity corresponding to 5nm depth into the samples [18]. The y axes are normalized by the Nb signal point-to-point for each coupon. We found no obvious impurities which explain the dramatically lower RRR, so we consider that other factors, such as grain size, may govern the RRR.

In Fig. 4, we observe that the low RRR samples have less H, which suggests that some impurity is trapping the free hydrogen. This aligns with the weakened $Q_0$ slope seen in the low RRR cavity. We see that N-doping increases H, and further studies needed to understand heightened NbH-signal. In Fig. 5, we observe that the low RRR EP and LTB samples have more C, but the N-doped samples do not follow this trend. While interesting, the C alone cannot explain the drastic difference in RRR. In Fig. 6, N diffuses similarly for low and high RRR. This aligns with their similar $R_{BCS}$. We also observe some N in bulk of low RRR EP and LTB which does not occur in the corresponding high RRR samples. In Fig. 7, O diffuses similarly for EP and LTB in their respective purities. The O profiles do not explain the difference in the LTB tests, suggesting another impurity is responsible for the different $R_{BCS}$.

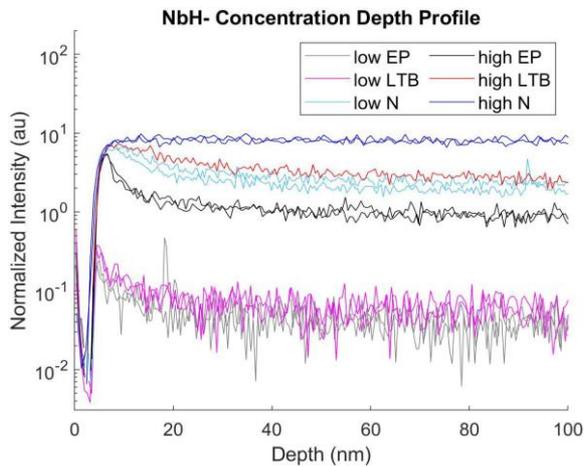

Figure 4: Impurity profile of NbH-.

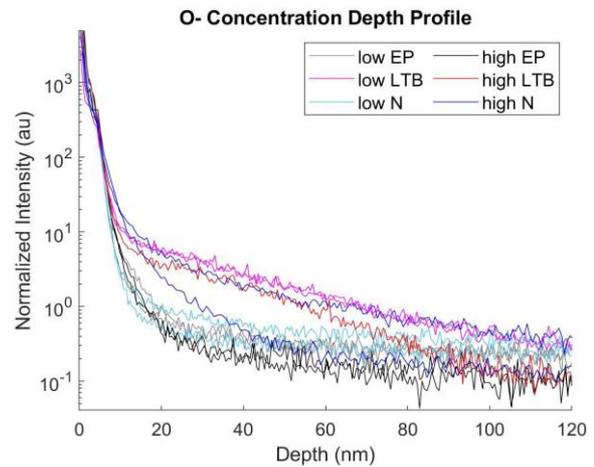

Figure 7: Impurity profile of O-.

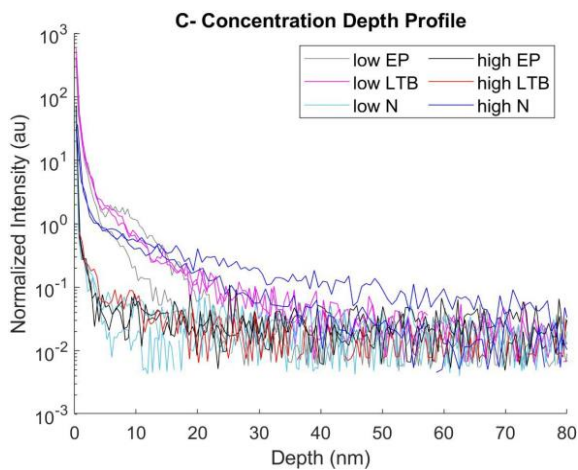

Figure 5: Impurity profile of C-.

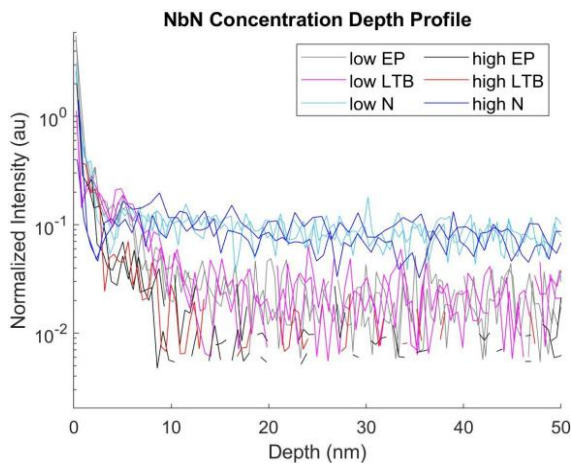

Figure 6: Impurity profile of NbN.

## CONCLUSION

The low RRR cavity behaves quite differently than high RRR cavities, with lower $R_{BCS}$, larger $R_{res}$, lower $Q_0$, and lower gradients in general. The intrinsic impurities affect the performance of the cavity for all surface treatments examined. Making the surface even dirtier allowed for lower $R_{BCS}$ even with a less clean bulk.

This difference is most notable in the EP testing, as the intrinsic impurities protect the cavity from a high field $Q_0$ slope and significantly improve the $R_{BCS}$. There is more similarity in the performance of the LTB cavities in terms of the offset of the $R_{res}$, the shape of the $R_{BCS}$ curves, and the O diffusion profiles. It is an important result that the combination of O and intrinsic impurities enables higher $Q_0$ and gradients. It appears that the LTB brought the low RRR cavity closer to the optimization of the $R_{BCS}$. The N-doping test showed increased $R_{res}$ from the other low RRR tests, but also showed a further decrease in the $R_{BCS}$. The similar diffusion of N, along with the similar $R_{BCS}$ shows that N-doping is a robust treatment in different purity SRF cavities. By understanding how O and N interact with the intrinsic impurities, we can gain insight how to develop a future high $Q_0$/high gradient surface treatment involving these impurities.